\documentclass[prd,preprint,superscriptaddress,amsmath,amssymb,nofootinbib]{revtex4}
\usepackage{graphicx}% Include figure files
\usepackage{dcolumn}% Align table columns on decimal point
\usepackage{bm}% bold math
\usepackage{amssymb}
\usepackage{amsmath}
\usepackage{epsfig}    
\usepackage{color}
\usepackage{slashed}
\usepackage{hhline}
\def\be{\begin{equation}}
\def\ee{\end{equation}}
\newcommand{\bea}{\begin{eqnarray}}
\newcommand{\eea}{\end{eqnarray}}
\newcommand{\nn}{\nonumber}

\begin{document}

{\begin{flushright}{KIAS-P20037, APCTP Pre2020 - 017}
\end{flushright}}

%%%%%%%%% Anomaly vs Excess; Which one would be preferred???
%candidate 1
%\title{Interpretation of XENON1T anomaly with a radiative seesaw model} 
%candidate 2
%\title{XENON1T anomaly makes a radiative seesaw model \\ "even more natural radiative seesaw model"?} 
%candidate 3
%\title{Even more natural radiative seesaw model \\via\\ an interpretation of XENON1T anomaly} 
%candidate 4
%\title{An attractive radiative neutrino model favored by XENON1T excess } 
%candidate 5
\title{A radiative seesaw model linking to XENON1T anomaly}

\author{Jongkuk Kim}
\email{jkkim@kias.re.kr}
\affiliation{School of Physics, KIAS, Seoul 02455, Korea}

\author{Takaaki Nomura}
\email{nomura@kias.re.kr}
\affiliation{School of Physics, KIAS, Seoul 02455, Korea}

\author{Hiroshi Okada}
\email{hiroshi.okada@apctp.org}
\affiliation{Asia Pacific Center for Theoretical Physics (APCTP) - Headquarters San 31, Hyoja-dong,
Nam-gu, Pohang 790-784, Korea}
\affiliation{Department of Physics, Pohang University of Science and Technology, Pohang 37673, Republic of Korea}

\date{\today}

\begin{abstract}
%We propose an interpretation of XENON1T anomaly with a radiative seesaw model 
% interpretation recently reported by XENON1t  is nicely explained by a radiative seesaw scenario. We 
We propose an attractive model that excess of electron recoil events around 1-5 keV reported by the XENON1T collaboration nicely links to the tiny neutrino masses based on a radiative seesaw scenario.
Our dark matter(DM) is an isospin singlet inert boson that plays an role in generating non-vanishing neutrino mass at one-loop level, and this DM inelastically interacts with a pair of electrons at one-loop level that is required to explain the XENON1T anomaly.
 It is also demanded that the mass difference between an excited DM and DM has to be of the order keV. 
%%%
Interestingly, the small mass difference $\sim$keV is proportional to the neutrino masses.
It suggests that we have double suppressions through the tiny mass difference and the one-loop effect. 
Then, we show some benchmark points to explain the XENON1T anomaly, satisfying all the constraints  
such as the event ratio of electrons of XENON1T, a long lived particle be longer than the age of Universe, and relic density in addition to the neutrino oscillation data and  lepton flavor violations(LFVs).

\end{abstract}
\maketitle
\newpage

\section{Introduction}
Dark matter (DM) is one of the important pieces to be understood its nature beyond the standard model (SM) and cosmology.
Recently, XENON1T collaboration reported an excess of electron recoil events around 1-5 keV energy over the known backgrounds~\cite{Aprile:2020tmw}.
After this report, a vast literature has arisen along this line of the subject such as explaining the excess by axions, absorption of keV scale DM, a scattering model, inelastic DM, boosted DM, and so on~\cite{Arcadi:2020zni, Choudhury:2020xui, Athron:2020maw, Shoemaker:2020kji, He:2020wjs, Davighi:2020vap, Okada:2020evk, Benakli:2020vng, Chigusa:2020bgq, Li:2020naa, Baek:2020owl, An:2020tcg, Ko:2020gdg, Gao:2020wfr, Cacciapaglia:2020kbf, Chao:2020yro, Ge:2020jfn, Bhattacherjee:2020qmv, DelleRose:2020pbh, McKeen:2020vpf, Chala:2020pbn, Bloch:2020uzh, Lindner:2020kko, Budnik:2020nwz, Gao:2020wer, Zu:2020idx, An:2020bxd, Bramante:2020zos, Nakayama:2020ikz, Primulando:2020rdk, Lee:2020wmh, Smirnov:2020zwf, Harigaya:2020ckz, Alonso-Alvarez:2020cdv,Fornal:2020npv, Boehm:2020ltd, Kannike:2020agf, Takahashi:2020bpq,Bally:2020yid,Su:2020zny,Du:2020ybt,DiLuzio:2020jjp,Bell:2020bes,Chen:2020gcl,Dey:2020sai,Buch:2020mrg,AristizabalSierra:2020edu,Choi:2020udy,Paz:2020pbc, 1807748, Ema:2020fit}.

In this paper, we propose a model that this excess by XENON1T marvelously links to tiny active neutrino masses based on a radiative seesaw scenario with a gauged hidden $U(1)$ symmetry~\cite{Nomura:2020azp}. Some of radiative seesaw models are renowned as natural models to connect  DM and the active neutrinos at low energy scale~\cite{Ma:2006km, Kajiyama:2013zla, Kajiyama:2013rla, Krauss:2002px, Aoki:2008av, Gustafsson:2012vj}.
More concretely, our DM is an isospin singlet inert boson that plays an role in generating non-vanishing neutrino mass at one-loop level, and this DM inelastically interacts with a pair of electrons at one-loop level that is required by the XENON1T anomaly.
It also demands that the mass difference between an excited DM and DM has to be of the order keV. 
%%%
The small mass difference $\sim$keV between DMs is proportional to the miniscule active neutrino masses. It suggests that we have double suppressions through this keV mass difference and the one-loop effect. 
At first, we show this mechanism and some benchmark points to explain the XENON1T anomaly, satisfying all the constraints  
such as the event ratio of electrons of XENON1T, a long lived particle be longer than the age of Universe, and relic density in addition to the neutrino oscillation data and LFVs. 

This paper is organized as follows.
In Sec.~II, we review our model, and construct our valid Lagrangian, Higgs potential, neutrino sector, LFVs, and $Z'$ boson mass.
In Sec.~III, we discuss our DM candidate to derive required scattering event rate with electrons, lifetime for a long lived particle, and cross sections of relic density. Then, we show our results accommodating all the issues discussed here. 
 In Sec.~IV, we devote to the summary of our results and the conclusion.

\section{Model}

\begin{table}[t!]
\begin{tabular}{|c||c|c||c|c||c|c|c|}\hline\hline  
& ~$L_L$~& ~$e_R$~  & ~$L'$~& ~$N$~& ~$H$~ & ~$\chi$~& ~$\varphi$~ \\\hline
%%%
$SU(2)_L$ & $\bm{2}$  & $\bm{1}$  & $\bm{1}$  & $\bm{1}$  & $\bm{2}$  & $\bm{1}$   & $\bm{1}$    \\\hline 
$U(1)_Y$   & $-\frac12$ & $-1$ & $-1$ & $0$  & $\frac12$  & $0$ & $0$  \\\hline
$U(1)_{X}$   & $0$ & $0$ & $1$   & $1$  & $0$  & $-1$  & $2$ \\\hline
\end{tabular}
\caption{ 
Charge assignments to fields in the model under $SU(2)_L\times U(1)_Y\times U(1)_X$ where all the new fields are neutral under $SU(3)_C$ and quark sector is exactly the same as the SM assignment.  }
\label{tab:1}
\end{table}

In this section, we briefly depict our model of local $U(1)_X$ dark sector.
We introduce three generations of isospin doublet and singlet vector-like fermions, which are respectively denoted by $L'\equiv[e',n']^T$ and $N$. 
%These connects to the neutrino masses.
 In scalar sector we add a source of DM candidate $\chi\equiv (\chi_R+i \chi_I)/\sqrt2$, and  
 $\varphi$ that plays a role in breaking gauged $U(1)_X$ symmetry spontaneously by developing its vacuum expectation value(VEV); $\langle \varphi \rangle = v_\varphi/\sqrt{2}$. Notice here that only the new fields have nonzero charges under this extra $U(1)$ symmetry, as can be seen in Table~\ref{tab:1}.
After the breaking of $U(1)_X$, we get a remnant  $Z_2$ symmetry that plays an role in assuring the stability of DM.
We explicitly write renormalizable valid Lagrangian under these symmetries as follows:     
\begin{align}
%& -{\cal L_M}  =   M_{L'} \bar L' L' + M_{N'} \bar N' N',  \label{Eq:Mass} \\
 -{\cal L_\ell}
&= y_{\ell} \bar L_L H e_R  +  f  \bar L_L  L'_R \chi  +  g \bar L'_L  \tilde H N_R + \tilde g \bar L'_R  \tilde H N_L  + h_L \bar N^c_L N_L \varphi + h_R \bar N^c_R N_R \varphi \nonumber\\
& +   M_{L'} \bar L' L' + M_{N} \bar N N + {\rm h.c.}, \label{Eq:yuk}
\end{align}
where $\tilde H = i \sigma_2 H^*$ $\sigma_2$ being second Pauli matrix, generation index is omitted, and $y_\ell$ is assume to be diagonal matrix without loss of generality due to the redefinitions of the fermions.
The scalar potential is also given by
\begin{align}
V = & \mu_H^2 H^\dagger H + \mu_\chi^2 \chi^* \chi + \mu_\varphi^2 \varphi^* \varphi + \mu (\chi^2 \varphi + c.c.) \nonumber \\
&  + \lambda_H (H^\dagger H)^2 + \lambda_{\varphi} (\varphi^* \varphi)^2 + \lambda_\chi (\chi^* \chi)^2 \nonumber \\
& + \lambda_{H \chi} (H^\dagger H)(\chi^* \chi) + \lambda_{H \varphi} (H^\dagger H)(\varphi^* \varphi) + \lambda_{\chi \varphi} (\chi^*\chi)(\varphi^* \varphi).
\end{align}
An important point of this potential is to have $\mu$ term that provides mass difference between the mass of real part and imaginary part  of $\chi$.
After symmetry breaking the squares of mass eigenvalues are explicitly written by
%\begin{equation}\frac{1}{2} \mu_\chi^2 (\chi_R^2 + \chi_I^2) + \frac{\mu v_\varphi}{ \sqrt{2}} (\chi_R^2 - \chi_I^2),\end{equation}which follows
{\begin{align}
& m_{\chi_R}^2 = m_\chi^2 + \frac{ \mu v_\varphi}{\sqrt{2}}, \quad
 m_{\chi_I}^2 = m_\chi^2 - \frac{ \mu v_\varphi}{\sqrt{2}}, \\
& m_\chi^2 = \mu_\chi^2 + \frac{\lambda_{H\chi}}{2} v^2  + \frac{\lambda_{\chi \phi}}{2} v_\phi^2,
\end{align}
where $v$ is VEV of Higgs field $H$.
Mass difference $\Delta m \equiv m_{\chi_R}-m_{\chi_I}$ is thus given by 
\begin{align}
\Delta m \sim \frac{  \mu v_\varphi}{ \sqrt{2} m_\chi},
\end{align}
where we assumed $\mu v_\varphi \ll m_\chi$.
The explanation of XENON1T anomaly by inelastic DM scattering requires ${\cal O}({\rm keV})$ of $\Delta m$ which will be analyzed in next section.
In addition, we show the neutrino mass matrix is proportional to $\Delta m$ below.}

\section{Dark matter}
 \begin{figure}[tb]
\includegraphics[width=80mm]{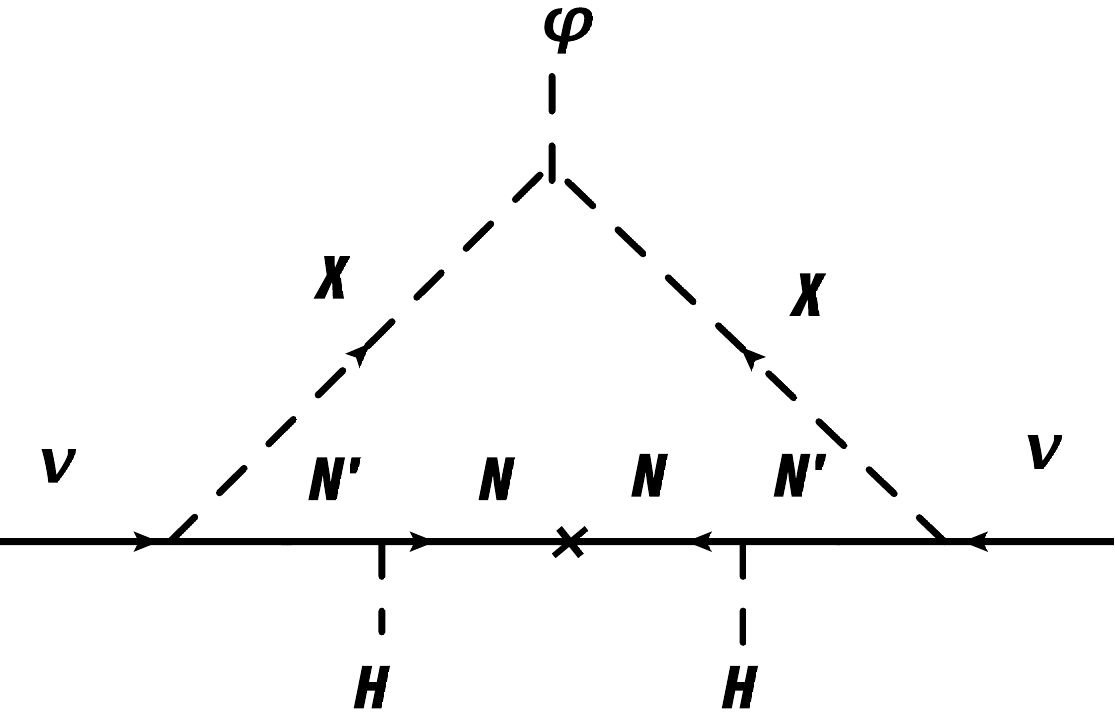}
\caption{Diagrams generating neutrino mass.}
\label{fig:neutrino}
\end{figure}

The neutrino mass matrix arises from the following Lagrangian 
\begin{equation}
L \supset \frac{1}{\sqrt{2}} f_{i a} V_{a \alpha} \bar \nu_{L_i} \psi^0_\alpha (\chi_R + i \chi_I) + h.c.,
\end{equation}
where $\psi$ is mass eigenstate of twelve extra neutral fermions, and $V$ is a unitary mixing matrix by twelve by twelve to diagonalize the extra neutral fermion matrix, as can be seen in Appendix \ref{app1}.
Then, we find the following neutrino mass matrix at one-loop diagram in Fig.~\ref{fig:neutrino}:
\begin{align}
\label{eq:nu_mass}
(m_\nu)_{ij} &=  \sum_{a, b=1}^3 \sum_{\alpha=1}^{12}
 \frac{f_{ia} V_{a \alpha} (f_{jb} V_{b \alpha})^T}{32 \pi^2} 
M_{\psi^0_\alpha}  \left[ \frac{m^2_{\chi_R}}{m_{\chi_R}^2 - M^2_{\psi^0_\alpha} } \ln \left( \frac{m^2_{\chi_R} }{M^2_{\psi^0_\alpha}} \right) 
  -  \frac{m^2_{\chi_I}}{m_{\chi_I}^2 - M^2_{\psi^0_\alpha} } \ln \left( \frac{m^2_{\chi_I} }{M^2_{\psi^0_\alpha}} \right)  \right]\nn\\
%%%
&\sim { \frac{m_\chi \Delta m }{16 \pi^2} }
\sum_{a, b=1}^3 \sum_{\alpha=1}^{12}
 \frac{f_{ia} V_{a \alpha}  (f_{jb} V_{b \alpha})^T M_{\psi^0_\alpha} }{m_\chi^2 - M_{\psi^0_\alpha}^2}
 \left[1 - \frac{M_{\psi^0_\alpha}^2}{m_\chi^2 - M^2_{\psi^0_\alpha} } \ln \left( \frac{m_\chi^2}{M^2_{\psi^0_\alpha}} \right)  \right],
\end{align}
where $m_\nu$ is diagonalized by a unitary matrix $V_{MNS}$ as $D_\nu = V_{MNS}^T m_\nu V_{MNS}$.
Applying Casas-Ibarra parametrization~\cite{Casas:2001sr} to our model, $f$ is rewritten in terms of the other parameters as follows:
\begin{align}
\label{Eq:Yukawa-result}
f&= V_{MNS}^* D_\nu^{1/2} V  {\cal O} (T^T)^{-1} ,
\end{align}
where ${\cal O}$ is three by three orthogonal matrix with arbitrary parameters, $T$ is an upper-right triangle matrix arisen from Cholesky decomposition. (See Appendix in ref.~\cite{Nomura:2016run}.)
%Hereafter, we use this $f$ to analyze the LFVs and DM analyses.

{\it $\ell_i \to \ell_j \gamma$ process}
The relevant lepton flavor violating(LFV) process arises from 
\begin{equation}
\label{Eq:intLFV}
f_{i a} \bar L_L^i L'^a_R S + h.c. \supset f_{i a} \bar \ell_L^i e'^a_R \chi +  f_{i a}^* \bar e'^a_R \ell_R^i\chi^*,
\end{equation}
where we consider $\chi$ as a complex scalar because of tiny mass difference.
Then, the branching rations(BRs) are given by
\begin{align}
&{\rm BR}(\ell_i\to\ell_j\gamma)\approx\frac{48\pi^3\alpha_{em}C_{ij}}{G_F^2 (4\pi)^4}
\left|\sum_{a} f_{j a} f^*_{i a} F(m_\chi,m_{E_a})\right|^2,\\
%%%
%%%
&F(m_a,m_b)\approx\frac{2 m^6_a+3m^4_am^2_b-6m^2_am^4_b+m^6_b+12m^4_am^2_b\ln\left(\frac{m_b}{m_a}\right)}{12(m^2_a-m^2_b)^4},
\end{align}
where $C_{21}=1$, $C_{31}=0.1784$, $C_{32}=0.1736$, $\alpha_{em}(m_Z)=1/128.9$, and $G_F$ is the Fermi constant $G_F=1.166\times10^{-5}$ GeV$^{-2}$.
The current experimental upper bounds are given by~\cite{TheMEG:2016wtm, Aubert:2009ag, Renga:2018fpd}
\begin{align}
{\rm BR}(\mu\to e\gamma)\lesssim 4.2\times10^{-13},\quad 
{\rm BR}(\tau\to e\gamma)\lesssim 3.3\times10^{-8},\quad
{\rm BR}(\tau\to\mu\gamma)\lesssim 4.4\times10^{-8}.
\label{eq:lfvs-cond}
\end{align}
%

\if0
A new contribution to the muon $g-2$, $\Delta a_\mu$ is also found in the same term as
\begin{equation}
\Delta a_\mu \simeq \frac{m_\mu^2}{8 \pi^2} \sum_{a} f_{2 a} f^*_{2 a} F(m_\chi,m_{E_a}),
\end{equation}
where $m_\mu$ is the muon mass.
\fi

\subsection{$Z'$ boson mass }

After the spontaneous symmetry breaking of $U(1)_X$ by the VEV of $\varphi$, we find the massive extra gauge boson $Z'$;
\begin{equation}
m_{Z'} = 2 g_X v_\varphi,
\end{equation}
where $g_X$ is the gauge coupling associated with $U(1)_X$.
Note also that we assume vanishing kinetic mixing between $U(1)_X$ and $U(1)_{em}$ and $U(1)_Y$.
%As we take $v_\varphi = \mathcal{O}(100)$ MeV, the mass of $Z'$ is $m_{Z'} \lesssim 100$ MeV in our scenario.

\section{Dark matter}
 \begin{figure}[tb]
\includegraphics[width=120mm]{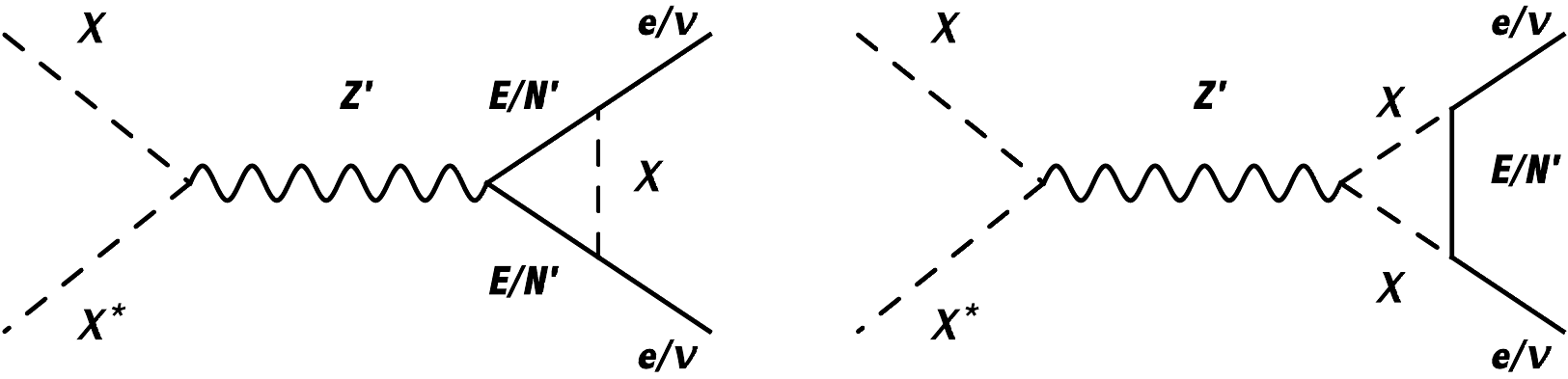}
\caption{Diagrams inducing effective Lagrangian for DM-lepton scattering containing one-loop effect.}
\label{fig:diagram-meZp}
\end{figure}

In this section, we discuss our dark matter explaining XENON1T anomaly via inelastic scattering.
We also discuss consistency of the model considering relic density, lifetime of excited DM and connection to neutrino mass.

\subsection{Couplings with electrons}
In order to fit the data of XENON1T excess, our DMs $\chi_{I,R}$ have to interact with electrons.
Such interaction is realized by $Z'$ extra gauge boson through one-loop diagram in Fig.~\ref{fig:diagram-meZp}. 
Here, we show relevant effective Lagrangian as follows:~\cite{Hutauruk:2019crc, Ko:2017yrd}
\begin{align}
\mathcal{L}_{eff} &= \delta g^\ell_{ij} \frac{g'}{m_Z'^2} [\bar\ell_j\gamma^\mu P_L \ell_i][\partial_\mu \chi_I \chi_R - \chi_I \partial_\mu \chi_R] 
+\delta g^\nu_{ij} \frac{g'}{m_Z'^2} [\bar\nu_j\gamma^\mu P_L \nu_i][\partial_\mu \chi_I \chi_R - \chi_I \partial_\mu \chi_R], \\
\delta g^\ell_{ij}&\simeq
\frac{g'}{(4\pi)^2} \sum_{a=1-3} f_{ia} f^\dag_{aj} 
\int[dx]_3 \ln\left(\frac{-z(x m_{\ell_i}^2 + y m_{\ell_j}^2)+x M_{E_a}^2 +(1-x) m^2_\chi}
{-z(x m_{\ell_i}^2 + y m_{\ell_j}^2)+x m_\chi^2 +(1-x) M_{E_a}^2}\right), \\
\delta g^\nu_{ij}&\simeq
\frac{g'}{(4\pi)^2} \sum_{a=1}^3 \sum_{b=1}^3 \sum_{\alpha=1}^{12} f_{ia}V_{a \alpha}  (f_{jb} V_{b \alpha} )^\dag
\int[dx]_3 \ln\left(\frac{x M_{\psi^0_\alpha}^2 +(1-x) m^2_\chi}
{x m_\chi^2 +(1-x) M_{\psi^0_\alpha}^2}\right),
%\frac{-M^4_{E_a} + m^4_\chi + 2 M^2_{E_a} m^2_\chi \ln\left[\frac{ M^2_{E_a}}{m^2_\chi}\right]}{(M^2_{E_a} - m^2_\chi)^2}
\end{align}
where $[dx]_3\equiv \int_0^1 dxdydz\delta(1-x-y-z)$.
%where we have assumed to be $m^2_{\ell_{i,j}}<<M_{E_a},m_\chi$.
Then, inelastic scattering cross section $\chi_I e\to \chi_R e$ is given by
\begin{align}
\sigma\approx \frac{g'^2 (\delta g^{\ell}_{11})^2 m_e^2 m_\chi^2}{4\pi m_{Z'}^4(m_e+m_\chi)^2}.
\end{align}
Applying this scattering cross section, we estimate the event rate $R$ as follows:
\begin{align}
 R\approx 1.13\times 10^{10}\times \left[\frac{(g' \delta g^\ell_{11})^2}{{\rm ton}\times{\rm year}}\right] 
\times\left[\frac{1{\rm GeV}}{m_\chi}\right] \times \left[ \frac{1 \ {\rm GeV}}{m_{Z'}} \right]^4,
\end{align}
where we have assumed to be $m_\chi<m_{Z'}$.
Here number of excess events observed by XENON1t is $N_{ex} \simeq 50$.

\subsection{Lifetime of excited DM}\label{subsec:lifetime}
In order to explain the anomaly of XENON1T, exited DM state should have long lifetime that is longer than the age of Universe ${\cal O}(10^{17})$ second at least. This must inevitably be considered since DM interacts with a pair of  neutrinos with the same coupling as a pair of electrons.
The lifetime of excited DM via the same process of neutrino scattering is estimated by~\footnote{This process would be bounded from SuperKamiokande that its lifetime be ${\cal O}(10^{24})$ second, but this bound is valid only for the case where the missing energy of neutrino is greater than ${\cal O}(0.1)$ GeV. Thus, this even stronger bound is not needed to be applied to our model.} 
\begin{align}
&\tau(\chi_R)=\frac{1}{\Gamma(\chi_R\to \chi_I \nu_i\bar\nu_j)},\\
&\Gamma(\chi_R\to \chi_I \nu_i\bar\nu_j)\approx
\frac{g'^2 N_\nu (\delta g^{\nu}_{ij})^2 \Delta m}{120\pi^3}\left(\frac{ \Delta m}{ m_{Z'}}\right)^4,
%+{\cal O}\left(\frac{ \Delta m}{ m_{Z'}}\right)^6,
\end{align}
where $N_\nu\sim3$ is the effective neutrino number, and the process of $\chi_R\to\chi_I e\bar e$ is kinematically forbidden due to $\Delta m^2\sim {\cal O}$(keV).
We can also neglect the other processes such as $\chi_R\to\chi_I 3\gamma$~\footnote{Notice here that $\chi_R\to\chi_I \gamma$ and $\chi_R\to\chi_I 2\gamma$ are forbidden or highly suppressed. } that is suppressed enough.
From this constraint, we find $f\lesssim \mathcal{O}(0.01)$--$\mathcal{O}(0.1)$ at $M_{L'}=100$ GeV depending on mass of $Z'$.

\subsection{Relic density of DM}
%{\color{blue} 
Since the mass difference between the two dark matters is very small, we can consider the DM as complex scalar field when calculating the relic density of DM. The corresponding DM relic density is obtained by 
\begin{eqnarray}
\Omega_{ \chi}h^2 &\simeq& \frac{1.6 \times 10^{-10} {\rm GeV^{-2}} }{\sqrt{g_*}\langle\sigma  v \rangle/x_f },
\end{eqnarray}
where $g_*$ is the effective number of relativistic degrees of freedom, and $x_f=m_{\chi}/T_f$. Here we take $\sqrt{g_*}\sim 3.8$ and $x_f\sim 10$ as inputs.
We have several processes to explain the relic density of DM, but we select one process of $\chi\chi^*\to \nu_i\nu_j/ee/\mu \mu$ via $Z'$ boson s-channel exchange. Since these processes are induced at one-loop level, we make use of the resonant effect around the pole at $2m_\chi\approx m_{Z'}$. In expansion in terms of relative velocity of DM, the cross section is approximately given by
\begin{align}
\sigma v_{\rm rel}(\chi\chi^*\to f\bar f)\approx \sum_{a=\ell,\nu}\sum_{i,j=1}^3\frac{g'^2(\delta g^{a}_{ij})^2 m_\chi^2}
{12\pi (m^4_{Z'}+16 m_\chi^4+m^2_{Z'}(\Gamma_{Z'}^2-8 m_\chi^2))} v^2_{\rm rel},
\end{align}
where $f$ are all the lepton flavors, kinematical condition is implicitly imposed, and we have assumed massless limit for final lepton masses for simplicity.
Notice here that this process can be evaded from the constraint of CMB because of p-wave dominant~\cite{Spergel:2003cb}.

Hereafter we mention the other processes, which is potentially taken into consideration.
The cross section via Yukawa coupling $f$  is induced by a process of $\chi\chi^*\to\ell\bar\ell(\nu\bar\nu)$ whose formula is given by $\frac{|f|^4}{48\pi m_\chi^2}\left(\frac{m_\chi}{M_{L'}}\right)^4$. This is suppressed by a factor $(m_\chi/M_{L'})^2\lesssim 10^{-4}$ where we fix $M_{L'}=100$ GeV and $m_\chi=1$ GeV, and $f\sim0.1$
that originates from the constraint that the lifetime of $\chi_R$ should be longer than the age of Universe, as discussed in Sec.~\ref{subsec:lifetime}. Therefore, the typical cross section is about $10^{-14}$ GeV$^{-2}$ at $m_\chi=1$ GeV.
This is very small compared to $10^{-9}$ GeV$^{-2}$ that gives 0.12 relic density of DM. 

In addition we kinematically forbid the process of $\chi\chi^*\to Z'Z'$ via t-channel by assuming $m_\chi \ll m_{Z'}$, since this process is s-wave dominant that gives stringent constraint from CMB.

Note also that the process of $\chi\chi^*\to \varphi Z'$ is basically allowed due to the p-wave dominant and might be found in an  appropriate cross section, but we here neglect this process for simplicity. We can always turn this cross section off, by assuming $m_\chi \ll m_\varphi,m_{Z'}$.

\subsection{Analysis}
\label{analysis}
%%%%%%%
 \begin{figure}[tb]
\includegraphics[width=80mm]{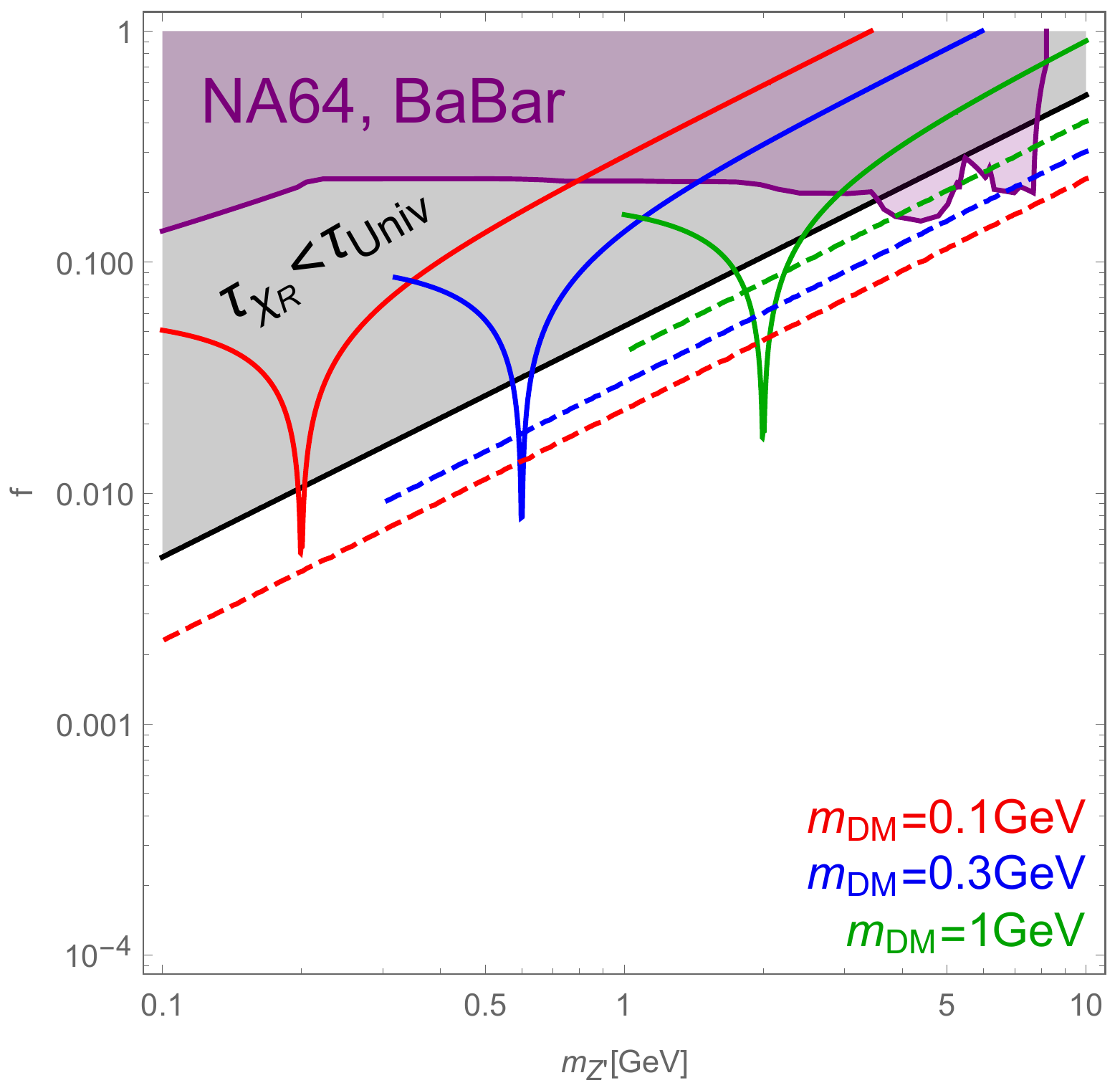}
\caption{Allowed line plots in terms of $m_{Z'}$ and $f$, where we fix $M_{L'}=100$ GeV
and $g'=\sqrt{4\pi}$.
Each of the red, blue, and green line represents the cross section to satisfy the relic density $\sim$0.12 at $m_\chi=0.1$ GeV, $0.3$ GeV, and $1$ GeV. While each of the dotted line shows the observed excess of events reported by XENON1t $N_{ex} \sim$50. The gray region is excluded since the lifetime of excited DM is shorter than the age of Universe, and the purple one  
is excluded by the NA64 and the BaBar experiments.}
\label{fig:result}
\end{figure}
%%%%%%%%
We show our result to accommodate all the constraints as we discussed above, where we also consider the constraint of BaBar~\cite{Lees:2014xha} and NA64~\cite{NA64:2019imj}.
Instead of global analysis, we express some benchmark points in order to clearly find tendency of our solution.
In Fig.~\ref{fig:result}, we demonstrate the allowed line plots in terms of $m_{Z'}$ and $f$, where we fix $M_{L'}=100$ GeV
and $g'=\sqrt{4\pi}$ that is perturbative limit; here we assume universal $f$ for illustration.
Each of the red, blue, and green line represents the cross section to satisfy the relic density $\sim$0.12 at $m_\chi=0.1$ GeV, $0.3$ GeV, and $1$ GeV. While each of the dotted line shows the observed excess of events  reported by XENON1T, $N_{ex} \sim$50. The gray region is excluded since the lifetime of excited DM is shorter than the age of Universe, and the purple one  
is excluded by the NA64 and the BaBar experiments. 
%%%
This figure suggests that the case of red line is marginally disfavored by the XENON1t anomaly,
but the other two cases of blue and green definitely have allowed points around $(m_{Z'},f)=(0.55\ {\rm GeV},0.012)$ and $(m_{Z'},f)=(2\ {\rm GeV},0.1)$, respectively~\footnote{In case of blue line, LFVs are totally safe because of the upper bound on Yukawa coupling is 0.01 at $M_{L'}=100$ GeV. On the other hand, the case of green can evade the constraints of LFVs by suppressing components of $f$ related to $\mu\to e\gamma$; $\sum_{a=1}^3f_{1a}f^*_{2a}\lesssim10^{-4}$.   }.
One can easily suppose that it would be difficult to explain the XENON1t anomaly above 1 GeV of the DM mass.

{
Finally, we briefly discuss consistency between the explanation of XENON1T anomaly and neutrino mass generation mechanism.
The scale of neutrino mass is roughly estimated from Eq.~\eqref{eq:nu_mass} as 
\begin{equation}
m_\nu \sim 10^{-9} {\rm [GeV]} \times \sum (f V)(fV)^T \frac{m_\chi}{\rm [GeV]} \frac{\Delta m}{2 \ {\rm [keV]}} \frac{100 \ {\rm [GeV]}}{M_{\psi^0}},
\end{equation}
where flavor index is omitted.
Thus Yukawa couplings $f$'s with $\mathcal{O}(0.01)$ to $\mathcal{O}(0.1)$ values are also suitable to generate neutrino mass since 
typical order of mass matrix elements is $\mathcal{O}(10^{-13})$ GeV to $\mathcal{O}(10^{-11})$ GeV.
Therefore our model can realize consistent framework linking the explanation of XENON1T anomaly and neutrino mass generation.
}

\section{Summary and Conclusion}
We have successfully explained excess of electron recoil events around 1-5 keV reported by the XENON1T collaboration
based on a radiative seesaw scenario. Our DM is an isospin singlet inert boson that plays an role in generating non-vanishing neutrino mass at one-loop level, and this DM inelastically interacts with a pair of electrons via one-loop level.
To explain the XENON1T anomaly, it demands that the mass difference between an excited DM and DM has to be of the order keV. 
%%%
Interestingly, we have shown this small mass difference is also proportional to the neutrino mass matrix.
Therefore, we have double suppressions to the active neutrino masses through the tiny mass difference $\sim$keV requested by the XENON1T anomaly as well as the one-loop effect. 
%It would be realized even more natural explanation for the tininess of neutrino masses.
We have displayed our valid processes to satisfy the event ratio of electrons of XENON1T, the lifetime of excited DM, and relic density in addition to the neutrino oscillation data and LFVs. 
Here, we have made the use of resonance effect around $2m_\chi\sim m_{Z'}$ to explain the relic density, and 
we have found a favored region around this point, as we have discussed in the last part of Section~\ref{analysis}.

%\newpage
%%%%%%%%%%%%%%%%%%%%%%%%%%%%%%%%%%%
\section*{Acknowledgments}
The work is supported in part by KIAS Individual Grants, Grant No. PG074201 (JK) and No. PG054702 (TN) at Korea Institute for Advanced Study.
This research was supported by an appointment to the JRG Program at the APCTP through the Science and Technology Promotion Fund and Lottery Fund of the Korean Government. This was also supported by the Korean Local Governments - Gyeongsangbuk-do Province and Pohang City (H.O.). 
H.O.~is sincerely grateful for all the KIAS members and Prof. Pyungwon Ko who especially provides me huge hospitality and fruitful discussion during my stay.
\appendix
\section{Appendix}
%%%%%%%%%%%%%%%%%%%%%%%%%%%%%%%%%%%
\label{app1}
%%%%%%%%%%%%%%%%%%%%%%%%%%%%%%%%%%%
{\it Extra fermion sector}:
In this appendix, we define the exotic fermion mass matrix.
The mass matrix of extra charged lepton $e'$ is given by
\begin{equation}
M_{L'} \bar L' L' \supset M_{L'} \bar e' e'.
\end{equation}
where $E$ does not mix with the SM charged leptons thanks to a remnant $Z_2$ symmetry.
The heavier neutral mass matrix, which is Majorana mass matrix, is found as
\begin{align}
L_{M_M} &=  \begin{pmatrix} \bar X^c_{1a} \\ \bar X^c_{2a} \\ \bar X^c_{3a} \\ \bar X^c_{4a} \end{pmatrix}^T 
\begin{pmatrix} 
{\bm 0}_{ab} & (M_{L'}^T)_{ab} & {\bm 0}_{ab} & (\tilde M_D^T)_{ab} \\ 
(M_{L'})_{ab} & {\bm 0}_{ab} &  (M_D)_{ab} & {\bm 0}_{ab} \\
{\bm 0}_{ab} & (M_D^T)_{ab} & (M_{N'_{LL}})_{ab} & (M_{N'}^T)_{ab} \\
 (\tilde M_D)_{ab} & {\bm 0}_{ab} &  (M_{N'})_{ab} & (M_{N'_{RR}})_{ab} 
\end{pmatrix}
\begin{pmatrix}  X_{1b} \\  X_{2b} \\  X_{3b} \\  X_{4b} \end{pmatrix} \nonumber \\
& \equiv \frac12 \bar X^c (M_X) X,
\end{align}
where $M_D = g v /\sqrt{2}$, $\tilde M_D = \tilde g v /\sqrt{2}$ and $M_{N'_{LL(RR)}} = h_{L(R)} v_\varphi/\sqrt{2}$.
We then rewrite fields by $n'_R \equiv X_1$, $n^{'c}_L \equiv X_2$, $N_R \equiv X_3$ and $N^c_L \equiv X_4$.
Then the mass matrix can be diagonalized by acting a unitary matrix as 
\begin{align}
V^T M_{X} V =D_N,  \quad X_{i a} =V_{ [a + 3 i -3] \alpha} \psi_\alpha^0, \label{eq:N-mix}
\end{align} 
where $\psi^0_\alpha$ is the mass eigenstate.
%\begin{align}- L_{M_N} = & (M_{L'} \bar N_L N_R + M_{N'} \bar N'_L N'_R  + M_D \bar N_L N'_R + \tilde M_D \bar N_R N'_L + h.c.  )
%\nonumber \\& + M_{N'_{LL}} \bar N'^c_L N'_L + M_{N'_{RR}} \bar N'^c_R N'_R,\end{align}

%\section{ Appendix}
%%%%%%%%%%%%%%%%%%%...

\end{document}